\documentclass[review]{elsarticle}

\usepackage{enumitem}

\usepackage{times}

\usepackage[T1]{fontenc}
\usepackage[scaled=0.81]{beramono}

\usepackage{hyperref}

\usepackage{graphicx}
\usepackage{epstopdf}
\usepackage{colortbl}

\usepackage{url}

\usepackage{listings}
\usepackage{eiffel}
\usepackage{boogie}
\usepackage{framed}
\usepackage{fancyvrb}
\usepackage{caption}

\usepackage[utf8]{inputenc} 
\usepackage[T1]{fontenc} 

\usepackage{lineno,hyperref}
\usepackage{framed}

\newcommand{\e}[1]{\mbox{\lstinline|#1|}}

\modulolinenumbers[200]

\begin{document}

\begin{frontmatter}

\title{Towards operational natural language.}

\author[iu,tls]{Alexandr Naumchev}
\ead{a.naumchev@innopolis.ru}
\address[iu]{Innopolis University, Innopolis, Russian Federation}
\address[tls]{Paul Sabatier University, Toulouse, France}

\begin{abstract}
The multiplicity of software projects' stakeholders and activities leads to the multiplicity of software specification views and thus creates the need to establish mutual consistency between them. The process of establishing such consistency is error-prone and requires adequate tool support. The present article introduces specogramming -- an approach that treats a modern object-oriented integrated development environment as a word processor. The approach turns the process of documenting initial specifications into a simplified form of programming and turns structured-natural-language specifications into runnable programs that yield multiple consistent-by-construction views, one of which is structured natural language.
\end{abstract}


\end{frontmatter}

\linenumbers

\section{Introduction} \label{sec:introduction}
The multiplicity of specification views leads to the following problems:

\begin{enumerate}
	\item The problem of producing the views and keeping them in sync.
	\item The precedence problem, when two views run out of sync.
	\item Reliance on potentially ambiguous structured natural language, when the views' precedence is not clear.
\end{enumerate}

The following development situation illustrates these problems. It also illustrates the specogramming approach itself in the upcoming sections. Consider a quality assurance (QA) engineer who relies on a unit test view and a developer that relies on a structured-natural-language view, such as user stories. When the QA engineer finds a bug, the QA vs. development conversation begins. The developer does not agree with the unit test used to uncover the bug, which leads to discussing the original user story. Examination of the user story reveals that either the development or the QA engineer has misunderstood the original requirement. The situation results in a waste of time, intellectual, and emotional energy.

The present article introduces specogramming -- the process of specifications programming. A specogram is an object-oriented (OO) program that looks like structured natural language. Specogramming treats integrated development environments (IDE's) as word processors and natural-language texts as programs. Running a specogram results in the generation of the necessary, consistent-by-construction, specification views.

Specogramming solves the three problems above:

\begin{enumerate}
	\item The problem of keeping the views in sync.\\
	Changes happen only in specograms, which consistently propagate the changes to all the necessary views. 
	\item The precedence problem, when two views run out of sync.\\
	Specograms always have the highest precedence. Running the associated specogram will remove the inconsistency.
	\item Reliance on potentially ambiguous structured natural language, when the views' precedence is not clear.\\
	In specogramming, a structured-natural-language view is a program. A stakeholder can run this program at any time and see what it means as applied to the views concerning the stakeholder.
\end{enumerate}

Specogramming reconsiders the features of OOP and the supporting IDE's in the following way:

\begin{itemize}
	\item In a qualified call ``\e{target.call}'', the ``\e{target}'' and the ``\e{call}'' represent two word combinations that can follow, in this order, in a natural-language statement.
	\item When a period symbol is entered, the IDE lists the features available on the target object in accordance with its static type. This is a standard feature of the modern OO IDE's. Specogramming treats the offered features as possible continuations of the specified phrase.
	\item Specogramming treats classes as vocabularies. A vocabulary, when instantiated and queried, yields another vocabulary object. The new vocabulary object contains queries with names that are grammatically consistent with the name of the query that yielded the object. The static typing of vocabularies guarantees the grammatical consistency. Properly typed vocabularies guarantee that the compiler will accept only grammatically correct, human-readable specograms.
\end{itemize}

Specogramming assumes continuous development of new operational vocabularies to keep up with the rapidly growing natural-language vocabularies used for specifying software. A GitHub repository \cite{Naumchev2017Specogram} contains several vocabularies and examples of their use that should be sufficient for developing the intuition behind the method. The project is currently being developed in Eiffel. Specogramming does not conceptually rely, however, on unique Eiffel's traits, and applies to any statically typed OO language. The article illustrates specogramming (\autoref{sec:specogramming}) on a specific example (\autoref{sec:the_gap}), describes the existing specogramming environment (\autoref{sec:specogramming_environment}), and concludes with an outline of the future work and recapitulates the method (\autoref{sec:conclusions_and_future_work}).

\section{The gap between structured-natural-language and formal specifications.} \label{sec:the_gap}

The present section formulates questions that motivate the invention of the specogramming approach.
Consider the following natural-language requirement, further referred to as ``requirement\_1'': ``a clock tick does not change the clock's hour if, in the beginning, the minute was smaller than 59''. A little bit more technical variant of this requirement expects basic knowledge of the OO notation from readers: ``a \e{clock.tick} does not change the \e{clock.hour} if, in the beginning, \e{clock.minute < 59}''.

The following parameterized unit test (PUT) \cite{Tillmann2005} exercises a candidate implementation of ``requirement\_1'':

\begin{lstlisting}
check_requirement_1 (c: CLOCK)
  require
    c.minute < 59
  do
    c.tick
  ensure
    c.hour ~ old c.hour
  end
\end{lstlisting}

Calling routine ``\e{check_requirement_1}'' with a specific \e{CLOCK} instance will test the implementation of the requirement if it meets the routine's precondition. The inability of the call to meet the precondition denotes irrelevance of the test with respect to the requirement, while the inability to pass the postcondition denotes a bug in the \e{CLOCK} implementation. This approach to testing through calling OO representation of ADT axioms is known as parameterized unit testing \cite{Tillmann2005}.

Adding a frame condition, such as ``\e{modify (clock)}'', to the routine's specification makes it usable as a driver for specifying the ``\e{tick}'' feature with a contract in the presence of a modular contract-based program prover \cite{Naumchev2016CompleteDrivers}. The following contract provably meets the ``\e{check_requirement_1}'' specification driver, which may be certified with AutoProof \cite{tschannen2015autoproof}, the prover of Eiffel programs:

\begin{lstlisting}
class CLOCK
  tick
    do
    ensure
      old minute < 59 implies hour ~ old hour
    end
end
\end{lstlisting}

The next task is to provide an implementation of ``\e{tick}'' that provably meets the specified contract. Program proving also makes it possible to use verification drivers for checking contracts' well-definedness \cite{Naumchev2016CompleteDrivers}. Because of the PUT's'/specification drivers' verifiability, both dynamic and static, the article uses it as the formal specification notation to illustrate specogramming.

The ``\e{check_requirement_1}'' PUT does not map to the original requirement, although it formally specifies its meaning. Namely, grasping the PUT requires the following, additional, knowledge of:
\begin{itemize}
	\item The Eiffel syntax.
	\item The notion of contract.
	\item The semantics of ``does not change'' as applied to contracts.
\end{itemize}

While Eiffel treats contracts as first-class citizens, other languages may not: .NET contracts, for example, look like ordinary instructions inside the routine's body, which further complicates grasping contracted .NET code.

These complications open the following questions. How to translate a structured-natural-language requirement to a verifiable form, such as PUT's, so that the translation process:
\begin{description}
\item [Q1] Hides details of a specific programming language (PL)?
\item [Q2] Hides the underlying contracts?
\item [Q3] Hides the detailed semantics of intuitively clear natural-language phrases, such as ``does not change'', ``increment'', ``decrement'', and many others?
\end{description}

Specogramming proposes a specific answer to these questions.

\section{Specogramming} \label{sec:specogramming}
The last modification of ``requirement\_1'' was: ``a \e{clock.tick} does not change the \e{clock.hour} if, in the beginning, \e{clock.minute < 59}''. Let us continue structuring it: ``execution\_of \e{"clock.tick"} does\_not\_change \e{"clock.hour"} if\_in\_the\_beginning \e{"clock.minute < 59"}''. This modification uses the underscore symbol to connect the words related to the requirement's structure, and quotes the domain-related terminology (the clock terminology in the ``requirement\_1'' example). The next iteration parenthesizes the problem domain-related terminology, and puts the period symbol after each closing parenthesis: 
\begin{lstlisting}
execution_of ("clock.tick").does_not_change ("clock.hour").
  if_in_the_beginning ("clock.minute < 59")
\end{lstlisting}
This form reflects the main idea behind specogramming: it treats structured natural language as object-oriented executable instructions. 

\begin{figure}
\begin{framed}
\begin{lstlisting}
requirement ("requirement_1").states_that_execution_of ("clock.tick").
  does_not_change ("clock.hour").for ("clock").of_type ({CLOCK}).
    if_in_the_beginning ("clock.minute < 59").period
\end{lstlisting}
\end{framed}
\caption{Application of specogramming to ``requirement\_1''}
\label{fig:requirement_1}
\end{figure}
The final form of the requirement adds something else (\autoref{fig:requirement_1}):
\begin{itemize}
\item The ``\e{requirement ("requirement_1")}'' call that labels the requirement for traceability.
\item The ``\e{.for ("clock").of_type (\{CLOCK\})}'' call adds the typing information about the ``\e{clock}'' variable. The ``\e{\{CLOCK\}}'' expression just yields string \e{"CLOCK"} in Eiffel. The advantage of this way of saying ``CLOCK'' is that the  compiler checks if the class exists or not.
\item The ``\e{.period}'' command call finalizes the whole instruction by yielding no object on which otherwise it would be possible to do more calls.
\end{itemize}
Each of the calls, except the ``\e{.period}'' call, yields an object. The static typing of these calls is such that the compiler does not accept structurally invalid requirements. If one forgets to add ``\e{.period}'' in the end, the compiler will remind that it is wrong to have a function call as the last call: one must finalize the instruction with a command call that yields nothing. The compiler cannot, however, rule out malformed inputs to the calls. The calls rule out malformed inputs at runtime, through preconditions: an attempt to run the same instruction as in \autoref{fig:requirement_1} but with ``\e{"requirement 1"}'' instead of ``\e{"requirement_1"}'' will fail: the precondition of the ``\e{requirement ()}'' function requires its input to be a well-formed identifier.

Compiling and running the specogram that contains the instruction in \autoref{fig:requirement_1} produces the following output:

\begin{itemize}
	\item A LaTeX document with an entry that turns into the following text when compiled to PDF:
		\begin{description}
			\item [requirement\_1:] Execution of $clock.tick$ does not change $clock.hour$ if, in the beginning, $clock.minute < 59$.
		\end{description}	
	The LaTeX entry resembles the original natural-language requirement with three parts (in \textit{italic}) formalized.
	\item A class with the following PUT:
	
	\begin{lstlisting}
	check_requirement_1
	-- execution of clock.tick does not change clock.hour
	-- if in the beginning clock.minute < 59 :
	-- for any
	    (clock: CLOCK)
	-- which
	  require
	-- that
	    clock.minute < 59
	  do
	-- executing
	    clock.tick
	-- will
	  ensure
	-- that
	    clock.hour ~ old clock.hour
	  end
	\end{lstlisting}

\end{itemize}

The specogram instruction in \autoref{fig:requirement_1} produces a PUT that not only contains the required code but also enriches it with human-readable information. The natural-language comments, that start with ``\e{--}'', make it possible to read the whole\\ ``\e{check_requirement_1}'' routine from the beginning to the end, as a holistic phrase. This is seamless approach \cite{knuth1984literate, walden1995seamless, Meyer:1997:OSC:261119, Meyer13Multi, Naumchev2017} that proposes to interweave the notations, not to switch between them. Hereafter the article uses term \textit{seamless requirement} \cite{Naumchev2017} to denote such readable-through routines, suitable for both software construction and verification, both dynamic and static.

%
%
%
%

The following procedure describes the process software development with specogramming as the software development methodology:
\begin{enumerate}
	\item Write a specogram in IDE.
	\item Compile the specogram and fix compilation errors, if any.
		\begin{enumerate}[label=\alph*]
			\item If a compilation error is caused by the inability of the compiler to recognize T in a ``\e{\{T\}}'' expression, declare the type.
		\end{enumerate}
	\item Run the resulting specogram.
	\item Compile seamless requirements resulting from the specogram at step 3.
	\item Repeatedly fix compilation errors detected at step 4., if any.
		\begin{enumerate}[label=\alph*]
			\item If an error talks about non-existing features or classes, create them.
			\item Go to step 1. and fix the specogram otherwise.
		\end{enumerate}
	\item Deploy a verification infrastructure for checking the resulting seamless requirements.
		\begin{enumerate}[label=\alph*]
			\item Call each of them with arguments that pass their preconditions, if you practice testing \cite{Tillmann2005}.
			\item Equip implementation classes with contracts that make the seamless requirements pass static verification if you use a static program prover \cite{Naumchev2016CompleteDrivers}.
		\end{enumerate}
	\item Provide an implementation that passes the checks from the verification infrastructure deployed at step 6.
		\begin{enumerate}[label=\alph*]
			\item Makes the calls from step 6.a. pass their respective seamless requirements' postconditions.
			\item Is provably correct against the contracts specified at step 6.b.
		\end{enumerate}			
\end{enumerate}

The implementation phase consists mainly of step 7., but it starts already at 2.a.: a successful compilation of a specogram assumes the existence of all types it talks about. Then, the implementation phase continues at step 5.a.: execution of the specogram turns a string expression of the form ``\e{target.call}'' into the actual call, and if the corresponding feature does not exist, the process requires to at least declare it. Step 5.b. assumes compilation errors caused by the initial specogram; non-declaration of a variable used in a specogram instruction is an example of such an error.

%

\section{Specogramming environment} \label{sec:specogramming_environment}
While the previous sections' purpose was bringing the intuition behind specogramming, the present section describes the specogramming environment.

The following example represents a complete specogram:

\begin{lstlisting}[numbers=left]
specify_software
  do
    create specification.further_referred_to_as ("clock_specification")
    requirement ("requirement_1").states_that_execution_of ("clock.tick").
      does_not_change ("clock.hour").for ("clock").of_type ({CLOCK}).
        if_in_the_beginning ("clock.minute < 59").period
    specification.writes_seamless requirements
    specification.writes_latex
  end
\end{lstlisting}
\begin{enumerate}
\item The instruction on line 3 instantiates a specification object.
\item The instruction spread across lines 4 to 6 adds a requirement object to the specification and specifies the requirement through the chain of qualified calls.
\item Line 7 writes the seamless requirements class. In this example, the class will contain only one seamless requirement, ``\e{check_requirement_1}'' (\autoref{sec:specogramming}).
\item Line 8 writes the latex document. The document will contain only one record in this case:
	\begin{description}
			\item [requirement\_1:] Execution of $clock.tick$ does not change $clock.hour$ if in the beginning $clock.minute < 59$.
	\end{description}
\end{enumerate}

The implementation of the ``\e{specify_software}'' routine is readable-through, from the beginning to the end, as a natural-language text.

\begin{figure}
	\centering
	\fbox{\includegraphics[width=0.95\linewidth]{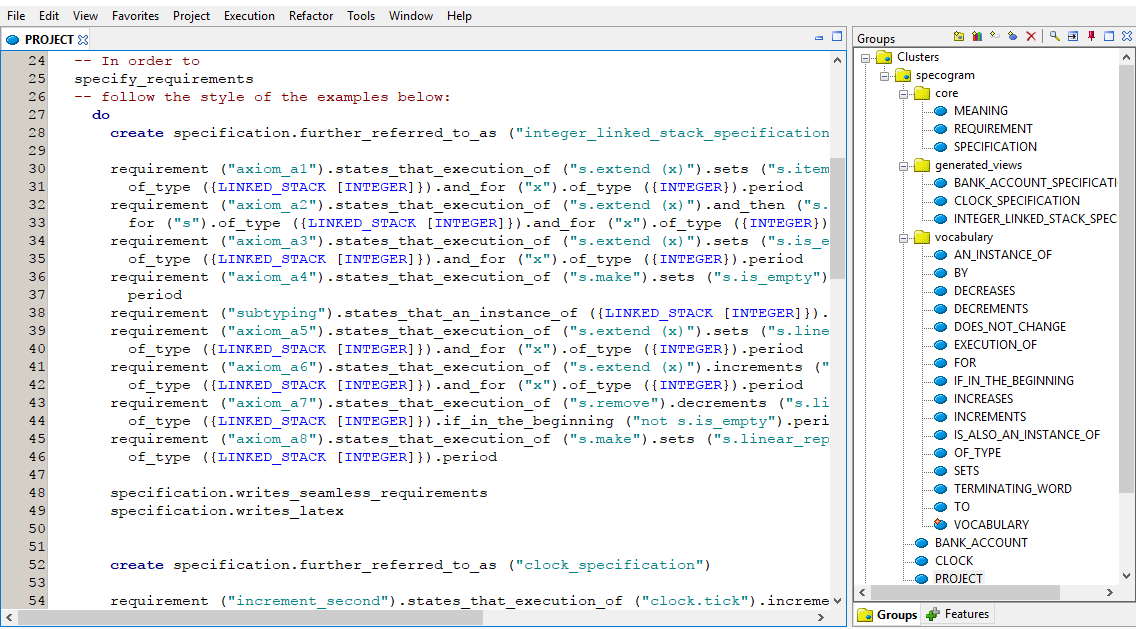}}
	\caption{EiffelStudio as a specogramming environment.}
	\label{fig:specogrampicture}
\end{figure}
A specogramming-based project may contain the following clusters (the rightmost pane, ``Groups'', in \autoref{fig:specogrampicture}):
\begin{description}
	\item[Core:] contains implementations of the most important requirements engineering concepts - \textit{requirement}, its hidden \textit{meaning}, and \textit{specification} that consists of requirements. This cluster is supposed to be changed when it is necessary to implement another view or improve existing views' generation.
	\item[Generated views:] stores specification views produced by specograms.
	\item[Vocabulary:] contains the vocabulary classes used for specogramming. This cluster is supposed to be modified every time a new meaningful vocabulary is found.
\end{description}

\section{Conclusions and future work} \label{sec:conclusions_and_future_work}

The idea of specogramming has a high potential. The current implementation of the natural-language-like vocabulary performs straightforward generation of structured specifications with elementary input checks. Nothing prevents enriching the implementation with advanced analysis of the requirements during specograms' execution. It is possible, in fact, to extend the existing vocabulary for producing and analyzing not only specifications but also implementations. In general, the vocabularies that look like natural language may hide intelligence of unlimited complexity.

Specogramming has the following immediate benefits to software specification practices:

\begin{itemize}
	\item Taking advantage of the OOP features, such as qualified calls, static typing, and command-query separation, to guarantee requirements' structural correctness.
	\item Taking advantage of modern IDE's' intelligent features, such as listing the services offered by an object, to facilitate the specification process.
	\item Taking advantage of the compiler for ruling out both malformed specograms and seamless requirements that they produce.
	\item Fixing a malformed view happens only in the original specogram, and rerunning it will propagate the fix to each concerned view.
	\item Wide applicability: all OO languages have qualified calls, the presence of which is the only assumption specogramming relies on.
\end{itemize}

To make the value of specogramming more evident, we need to do a considerable amount of work:

\begin{itemize}
	\item Develop methodological recommendations for developing vocabularies; investigate how much this development may be automated.
	\item Evaluate the approach on a meaningful example (such as Tokeneer \cite{Barnes2006}).
	\item Enrich the existing vocabulary with as many meaningful expressions as possible.
	\item Refine the current design of the solution which may be suboptimal.
	\item Possibly add traditional contracts to the current set of verifiable specifications produced by specograms.
\end{itemize}

The Specogram GitHub repository \cite{Naumchev2017Specogram} includes several examples, including the one used in the present article. Development of this project is happening inside this repository.


\bibliography{../library}

\end{document}